\documentclass{PoS}
\usepackage{graphicx}
\usepackage{indentfirst}
\usepackage{slashed}
\usepackage{amssymb}
\usepackage{amsmath}
\usepackage{bbm}
\usepackage{epsfig} 
\usepackage{multirow}%
\usepackage{epstopdf}
\usepackage{extarrows}
\usepackage{booktabs,lipsum,calc}

\newsavebox{\leftbox}
\newsavebox{\rightbox}

\newcommand{\be}{\begin{equation}} \newcommand{\ee}{\end{equation}}
\newcommand{\ba}{\begin{array}{c}} \newcommand{\ea}{\end{array}}
\newcommand{\bea}{\begin{eqnarray}} \newcommand{\eea}{\end{eqnarray}}

\title{Scalar form factors of semi-leptonic $D\to\pi/ \bar{K}$ transitions with coupled-channel effects}

\ShortTitle{Scalar form factors of semi-leptonic $D\to\pi/ \bar{K}$ transitions with coupled-channel effects}

\author{\speaker{De-Liang Yao}%
      \\
Instituto de F\'{\i}sica Corpuscular (centro mixto CSIC-UV), Institutos de Investigaci\'on de Paterna,
Apartado 22085, 46071, Valencia, Spain\\
      E-mail: \email{deliang.yao@ific.uv.es}}

\author{Miguel~Albaladejo\\
Departamento de F\'{\i}sica, Universidad de Murcia, E-30071 Murcia, Spain\\
E-mail: \email{miguel.albaladejo@ific.uv.es}}

\author{Pedro~Fern\'andez-Soler\\
Instituto de F\'{\i}sica Corpuscular (centro mixto CSIC-UV), Institutos de Investigaci\'on de Paterna,
Apartado 22085, 46071, Valencia, Spain\\
Email: \email{pedro.fernandez@ific.uv.es}}

\author{Feng-Kun Guo\\
CAS Key Laboratory of Theoretical Physics,  Institute of
Theoretical Physics, Chinese~Academy~of~Sciences,
Beijing~100190, China\\
School of Physical Sciences, University of Chinese Academy of
Sciences, Beijing 100049, China\\
Email: \email{fkguo@itp.ac.cn}}

\author{Juan~Nieves\\
Instituto de F\'{\i}sica Corpuscular (centro mixto CSIC-UV), Institutos de Investigaci\'on de Paterna,
Apartado 22085, 46071, Valencia, Spain\\
Email: \email{jmnieves@ific.uv.es}}

\abstract{Coupled-channel effects are taken into account for the study of scalar form factors in semi-leptonic $D\to\pi\bar{\ell}\nu_\ell$ and $D\to \bar{K}\bar{\ell}\nu_\ell$ decays, by solving the Muskhelishvili-Omn\`es integral equations.  As inputs, we employ the unitarized amplitudes taken from chiral effective theory for the region not far away from thresholds, while, at higher energies of the Goldstone bosons,  proper asymptotic conditions are employed. Within Muskhelishvili-Omn\`es formalism, the scalar form factors are represented in terms of Omn\`es matrix multiplied by a vector of polynomials. We reduce the number of subtraction constants by matching to the scalar form factors derived in chiral perturbation theory up to next-to-leading order. The recent lattice QCD data by ETM collaboration for ${D\to\pi}$ and ${D\to\bar{K}}$ scalar form factors are simultaneously well described. The scalar form factors for $D\to\eta$, $D_s\to \bar{K}$ and $D_s\to \eta$ transitions are predicted in their whole kinematical regions. Using our fitting parameters, we also extract the following Cabibbo--Kobayashi--Maskawa elements: $|V_{cd}|=0.243(_{-12}^{+11})_{\rm sta.}({3})_{\rm sys.}(3)_{\rm exp.}$ and $|V_{cs}|=0.950(_{-40}^{+39})_{\rm sta.}(1)_{\rm sys.}(7)_{\rm exp.}$. The approach used in this work can be straightforwardly extended to the semileptonic $B$ decays.}

\FullConference{XVII International Conference on Hadron Spectroscopy and Structure\\
		 25-29 September, 2017\\
		 University of Salamanca, Salamanca, Spain}

\begin{document}

\section{Introduction}
\begin{figure*}[t]
\begin{center}
\epsfig{file=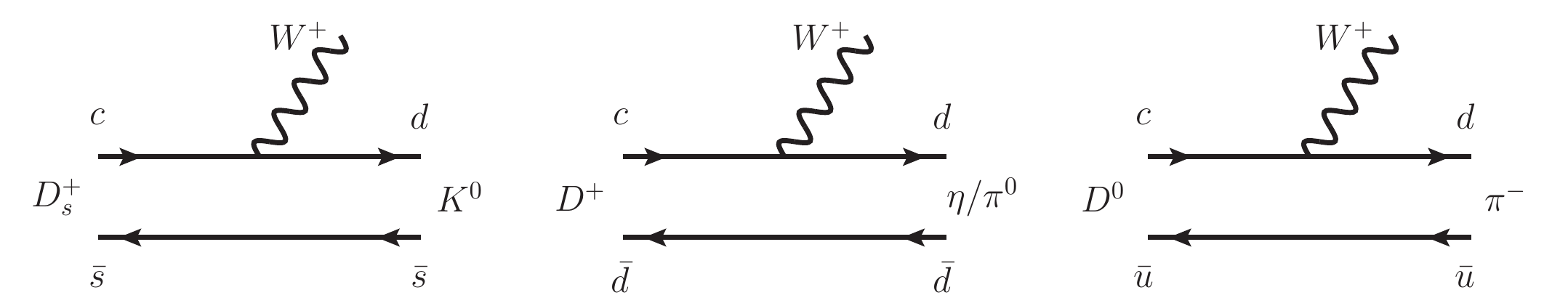,width=0.82\textwidth}
\epsfig{file=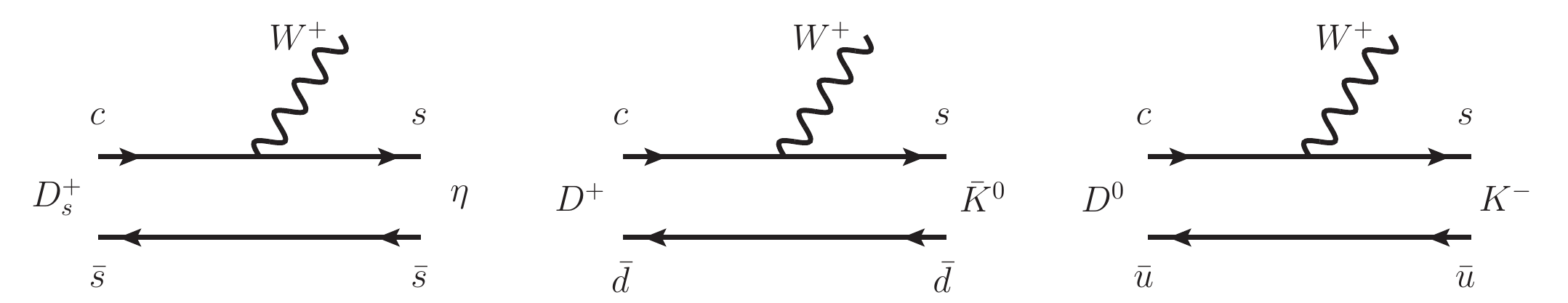,width=0.82\textwidth}
\caption{Semileptonic decays induced by the $c\to d$ and $c\to s$ flavour-changing transitions. \label{fig:hl}}
\end{center}
\end{figure*}

Semileptonic decays play a particularly important role in the precise determination of the Cabibbo--Kobayashi--Maskawa (CKM) elements, see Refs.~\cite{Richman:1995wm,Dingfelder:2016twb} for a review.
For a semileptonic decay of the type $P(p)\to{\phi}(p^\prime)\,{\ell}(p_\ell)\,\nu_\ell(p_\nu)$, the invariant amplitude reads
\bea
\mathcal{A}=\frac{G_FV_{qc}}{\sqrt{2}}\big\{\bar{u}(p_\ell)\gamma^\mu(1-\gamma_5)v(p_\nu)\big\}\big\{\langle{\phi}(p^\prime)|\bar{q}\gamma_\mu(1-\gamma_5)c|P(p)\rangle\big\}\ ,
\eea
where $G_F$ is the Fermi constant; $V_{qc}$ is the element of CKM matrix; $P\in \{D,D_s\}$ and $\phi\in\{\pi,K,\bar{K},\eta\}$. We will study the decays induced by the $c\to d$ and $c\to s$ flavour-changing transitions, depicted in Fig.~\ref{fig:hl}. The term in the first bracket is the weak matrix element, while the one in the second bracket corresponds to the hadronic part, which can be parametrized as
\bea
\langle{\phi}(p^\prime)|\bar{q}\gamma^\mu{c}|P(p)\rangle=f_+(q^2)\big[\Sigma^{\mu}-\frac{m_P^2-m_\phi^2}{q^2}q^\mu\big]+f_0(q^2)\frac{m_P^2-m_\phi^2}{q^2}q^\mu ,
\eea
with $\Sigma^\mu=(p+p^\prime)^\mu$ and $q^\mu=p^{\prime\mu}-p^\mu$.  The axial-vector part vanishes in accordance with parity conservation. Here $f_+(q^2)$ and $f_0(q^2)$ are called vector and scalar form factors, respectively, fulfilling the kinematical constraint $f_+(0)=f_0(0)$ at  $q^2=0$. In isospin basis, two multiplets of scalar form factors with definite strangeness and isospin, labeled by $(S,I)$, can be constructed:
\bea
\vec{\cal F}^{(0,\frac{1}{2})}(s)\overset{c\to d}{\equiv}\left(
\begin{array}{c}
  -\sqrt{\frac{3}{2}}  f_0^{D^0\to \pi^-}(s) \\
-f_0^{D^+\to{\eta}}(s)  \\
-f_0^{D_s^+\to{K^0}}(s)
\end{array}
\right)\ , \qquad 
\vec{\cal F}^{(1,0)}(s)\overset{c\to s}{\equiv}
\left(
\begin{array}{c}
  -\sqrt{2}  f_0^{D^0\to K^-}(s) \\
f_0^{D_s^+\to{\eta}}(s) 
\end{array}
\right)\ ,
\eea
with $s\equiv q^2$. Hereafter, for simplicity we will use the following notations: $f_0^{D\to \pi}=f_0^{D^0\to \pi^-}$, $f_0^{D\to \eta}=f_0^{D^+\to \eta}$, $f_0^{D_s\to K}=f_0^{D_s^+\to K^0}$, $f_0^{D\to \bar{K}}=f_0^{D^0\to K^-}$ and $f_0^{D_s\to \eta}=f_0^{D_s^+\to \eta}$. In the single channel case, dispersive analyses of the heavy-to-light scalar form factors were made in Refs.~\cite{Burdman:1996kr,Flynn:2000gd,Flynn:2006vr, Flynn:2007ki}. Here we intend to extend the study to the coupled-channel case by using the Muskhelishvili-Omn\`es (MO) formalism~\cite{Omnes:1958hv,Muskhelishvili}. The MO formalism has been widely applied to the scalar
$\pi\pi$, $\pi K$ and $\pi\eta$ form factors, e.g.,
Refs~\cite{Donoghue:1990xh,Liu:2000ff,Jamin:2001zq,Albaladejo:2015aca}.

\section{Muskhelishvili-Omn\`es representation}
\begin{figure*}[t]
\begin{center}
\epsfig{file=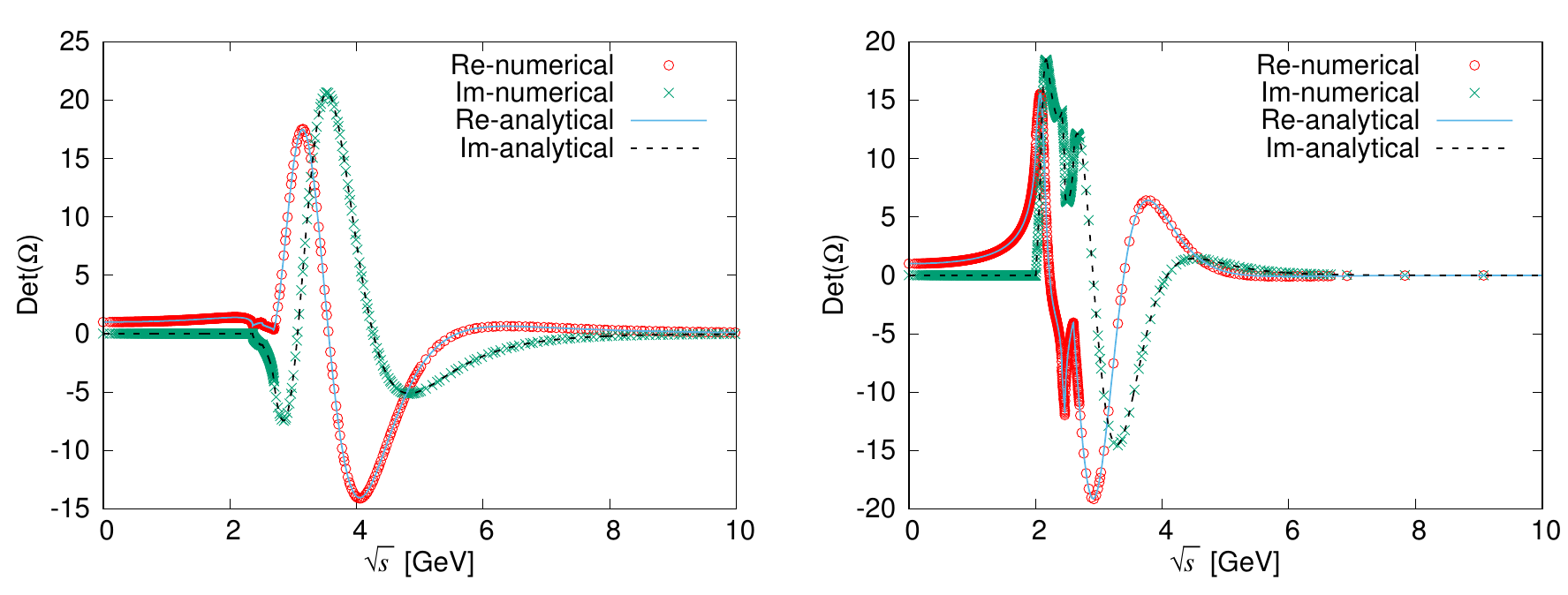,width=\textwidth}
\caption{Determinant of the MO matrices. Left: (S,I)=(1,0); Right: (S,I)=(0,1/2). The matching points are fixed at $\sqrt{s_m}=2.7$~GeV and $\sqrt{s_m}=2.6$~GeV, respectively. \label{fig:check}}
\end{center}
\end{figure*}

The discontinuity of the scalar form factor along the unitary cut reads (indices suppressed)
\bea\label{eq:uni}
{\rm Im}\vec{\cal F}(s)=\mathbf{T}^\ast(s)\Sigma(s)\vec{\cal F}(s)\ ,
\eea
where $\mathbf{T}$ is the $n$-coupled channel $P\phi$ amplitude in $S$-wave and $\Sigma(s)={\rm diag}\{\sigma_{P_1\phi_1},\cdots,\sigma_{P_n\phi_n}\}$ with
$
\sigma_{P_i\phi_i}=\sqrt{[s-(m_{P_i}+M_{\phi_i})^2][s-(m_{P_i}-M_{\phi_i})^2]}/s$.
It has an algebraic solution~\cite{Omnes:1958hv,Muskhelishvili}:
\bea\label{eq:representation}
\vec{\cal F}(s)= {\Omega}(s)\cdot \vec{\mathcal{P}}(s)\ ,
\eea
with $\Omega(s)$ the MO matrix and $\vec{\mathcal{P}}(s)$ a vector of polynomial components with real coeffecients. The MO matrix satisfies a un-subtracted dispersion relation,
\bea\label{eq:MO}
\Omega(s)=\frac{1}{\pi}\int_{s_{\rm th}}^{\infty}\frac{\mathbf{T}^\ast(s)\Sigma(s)\Omega(s)}{s^\prime-s-i\epsilon}{\rm d}s^\prime\ , \quad(s_{\rm th}:~\text{lowest threshold})
\eea
The above integral equation can be solved numerically using $\mathbf{T}$ as input. To that end, we take $\mathbf{T}=-\mathbf{T}^U/(16\pi^2)$ up to a energy point $s_m$, below which the unitarized amplitude $\mathbf{T}^U$ is valid. $\mathbf{T}^U$ is built from relativistic chiral potentials given, e.g., in Refs.~\cite{Liu:2012zya,Yao:2015qia,Du:2017ttu}.  All the relevant low-energy constants (LECs) and subtraction parameters were fixed by fitting to lattice QCD data of $S$-wave scattering lengths~\cite{Liu:2008rza,Liu:2012zya,Mohler:2013rwa,Lang:2014yfa}. Here we employ the next-to-leading order (NLO) potentials given in Ref.~\cite{Liu:2012zya} because the needed LECs are better determined than those that appear at one-loop level as discussed in Ref.~\cite{Du:2016tgp}. Furthermore, as shown in Ref.~\cite{Albaladejo:2016lbb}, the energy levels in the $(S,I)=(0,1/2)$ sector calculated based on the NLO potentials are in a good agreement with the lattice results of Ref.~\cite{Moir:2016srx}. Above $s_m$, phase shifts and inelasticities are extrapolated to the following values at infinity:
\bea\label{eq:inftyvalue}
\delta_i(\infty)=n\pi\,\delta_{i1}\ ,\quad \eta_i(\infty)=1\ ,\quad (i=1,\cdots,n)\ ,
\eea
by using the interpolating functions~\cite{Moussallam:1999aq}
$x_i(s)=x_i(\infty)+\frac{2[x_i^U(s_m)-x_i(\infty)]}{1+
({s}/{s_m})^{3/2}}$, $x\in\{\delta,\eta\}
$.  The asymptotic values in Eq.~\eqref{eq:inftyvalue} ensure that 
\bea
\lim_{s\to \infty}|\mathbf{T}_{ij}|=0\quad \text{for}\quad i\neq j \quad\text{and} \quad \lim_{s\to \infty}\sum_{i=1}^n\delta_i(s)=n\pi\ ,
\eea
which further guarantee that the MO equation~\eqref{eq:MO} has a unique solution, albeit a global normalization. For $n$-coupled channels with $n\leq3$, $\mathbf{T}$-matrix can be readily constructed from $\delta_i$ and $\eta_i$, as demonstrated in Refs.~\cite{Waldenstrom:1974zc,Lesniak:1996qx}.
It should be stressed that the determinant of the MO matrix also satisfies an Omn\`es-type dispersion relation~\cite{Omnes:1958hv,Muskhelishvili}
\bea
{\rm det}\, \Omega(s)=\exp\bigg[\frac{s}{\pi}\int_{s_{\rm th}}^\infty\frac{\psi(s^\prime)}{s^\prime(s^\prime-s-i\epsilon)}{\rm d}s^\prime\bigg]\ , 
\eea
where $\psi$ is determined from $\exp(2i\psi(s))\equiv {\rm det} [\mathbbm{1}+2i\mathbf{T}(s)\Sigma(s)]$. It provides a way to check the numerical solution of the MO integral equation. In Fig.~\ref{fig:check}, the determinants of the MO solutions are compared with their corresponding analytical results and excellent agreements are observed. 

The polynomials in Eq.~\eqref{eq:representation} can be constrained by matching the MO representation to the chiral results of the form factors at low Goldstone-boson energies. Up to $\mathcal{O}(s)$, one has
\bea\label{eq:pol}
\vec{\mathcal{P}}(s)=\vec{\cal F}_{\chi}(s_{m}^\chi)+\bigg\{(s-s_{m}^\chi)\bigg[{{\vec{\cal F}}^\prime_{\chi}}(s_m^\chi)-{\Omega}^\prime(s_m^\chi)\cdot \vec{\cal F}_{\chi}(s_{m}^\chi)\bigg]\bigg\}+\mathcal{O}(s^2)\ ,
\eea
where the MO matrix is chosen to be normalized to one at the chiral matching point $s_m^\chi$, i.e., $\Omega(s_m^\chi)=\mathbbm{1}$. The components of the chiral amplitude $\vec{\cal F}_{\chi}$ are given by
\bea
(S,I)=(0,\frac{1}{2})&:& \quad  {\cal F}_{\chi}^1(s)=\sqrt{\frac{3}{2}}f_{\chi}^{D\pi}(s)\ ,\quad {\cal F}_{\chi}^2(s)=\frac{1}{\sqrt{6}}f_{\chi}^{D\eta}(s)\ ,\quad {\cal F}_{\chi}^3(s)=f_{\chi}^{D_s\bar{K}}(s)\ ;\nonumber\\
(S,I)=(1,0)&:& \quad  {\cal F}_{\chi}^1(s)=\sqrt{2}f_{\chi}^{DK}(s)\ ,\quad {\cal F}_{\chi}^2(s)=\sqrt{\frac{2}{3}}f_{\chi}^{D_s\eta}(s)\ .
\eea
Above, the common function $f_{\chi}^{P\phi}(s)$ takes the explicit form~\cite{Wise:1992hn,Burdman:1992gh}
\bea\label{eq:chiral}
f_{\chi}^{P\phi}(s)&=&\frac{-1}
{\sqrt{2}F_0}\bigg\{\big[\frac{\tilde{g}\,\mathring{m}f_P}{m_*^{2}}+\beta_1\big]
\frac{m_P^2-M_\phi^2-s}{m_P^2-M_\phi^2}
+\big[f_P-{\beta_2}(m_P^2+M_\phi^2-s)\big]\frac{m_P^2-M_\phi^2+s}{2(m_P^2-M_\phi^2)}
\bigg\},\label{eq:f0chpt}
\eea
where $f_P$ and $\mathring{m}$ are the decay constant and mass of the charmed mesons in the chiral limit, respectively. Furthermore, $F_0$ is the Goldstone-boson decay constant in the chiral limit, and ${m_\ast}$ is the mass of the exchanged $P^\ast$ meson. $\tilde{g}$ is the coupling constant for the $PP^\ast\phi$ vertex.

\begin{table}[hbt]
\caption{Pole parameters of $D_{s0}^\ast(2317)$ \label{tab:pole}}
\centering
\begin{tabular}{ccc}
\hline
&$g_{DK}$\;[GeV]& $g_{D_s\eta}$\;[GeV]\\
$\sqrt{s_{p}}$\;[MeV]  & $|{\rm Residue}|^{1/2}$\;($DK$) &  $|{\rm Residue}|^{1/2}$\;($D_s\eta$) \\
\hline
$2315.2^{+18.4}_{-28.2}$ 			& $9.5^{+1.2}_{-1.1}$	 	& $7.5^{+0.5}_{-0.5}$ \\
\hline
\end{tabular}
\end{table}

To finish this section, it should be emphasized that the MO representation only accounts for the contribution corresponding to the unitary cut due to the rescattering of the $P\phi$ system. In the $(1,0)$ sector, in order to incorporate the contribution from the $D_{s0}^\ast(2317)$ state~\cite{Aubert:2003fg,Krokovny:2003zq} {\it below} the unitary cut, the following 
substitution should be employed,
\bea
\Omega^{(1,0)}\cdot\vec{\cal P}^{(1,0)}(s)\to\Omega^{(1,0)}\cdot\bigg\{\frac{\beta_0\,\vec{\Gamma}}{s-s_p}+\vec{\cal P}^{(1,0)}(s)\bigg\},
\eea
where $\beta_0$ is a free parameter and $\vec{\Gamma}=(g_{DK},g_{D_s\eta})^T$. The pole parameters $s_p$, $g_{DK}$ and $g_{D_s\eta}$ can be determined from the unitarized amplitude $\mathbf{T}^U$~\cite{Liu:2012zya,Guo:2015dha} and are compiled in Table~\ref{tab:pole}.

\section{Numerical results}
\begin{figure*}[t]
\begin{center}
\epsfig{file=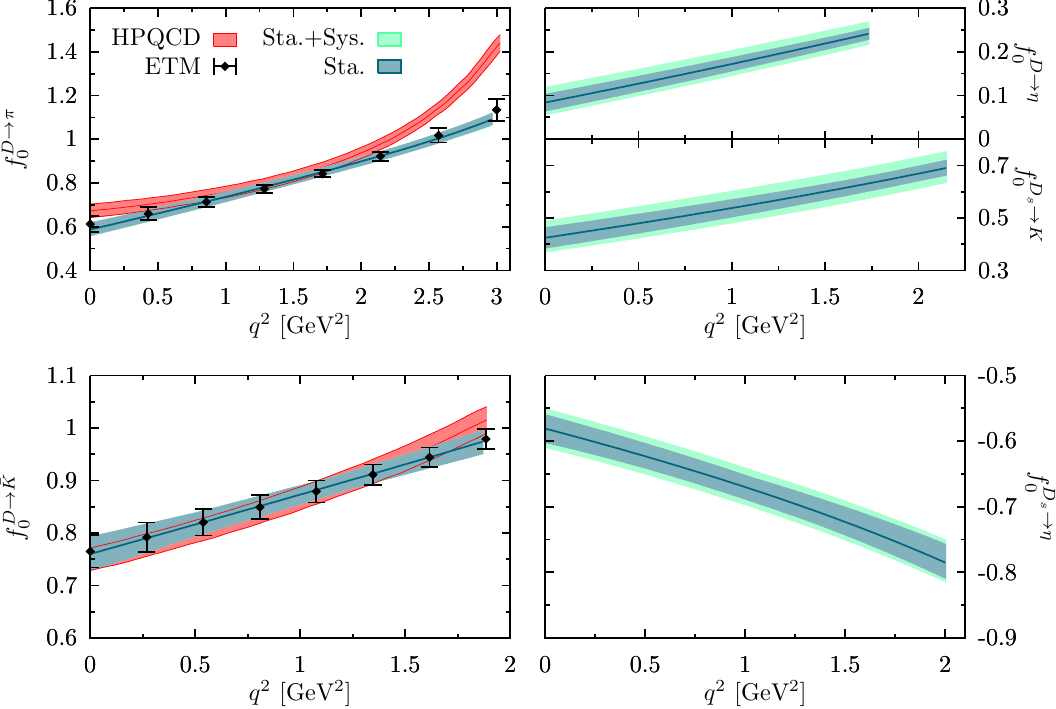,width=0.85\textwidth}
\caption{Scalar form factors in the $P\to\phi\bar{\ell}\nu_\ell$ decays. \label{fig:charmfit}}
\end{center}
\end{figure*}

Lattice QCD (LQCD) simulations for the $D\to\pi$ and $D\to \bar{K}$ scalar form factors were performed by the HPQCD collaboration~\cite{Na:2010uf,Na:2011mc} and very recently by the ETM collaboration~\cite{Lubicz:2017syv}. Due to the inclusion of hypercubic discretization effects~\cite{deSoto:2007ht},  the ETM results for the $D\to\pi$ form factor in the continuum limit are  significantly different from the HPQCD ones in the region close to $q_{\rm
max}^2=(m_D-M_\pi)^2$, whereas the changes for the $D\to \bar{K}$ form factor are less important. Therefore, we prefer to fit to the most recent ETM data only.\footnote{The covariance matrix, provided by ETM collaboration~\cite{Lubicz:2017syv} via private communication, is highly singular, hence we follow the approach of diagonal approximation, see, e.g., Ref.~\cite{Jang:2011fp} for more details.}

The relevant parameters in the MO representation appear in the chiral representation of the form factor in Eq.~\eqref{eq:chiral}. In our fitting procedure, we set $\tilde{g}=(1.133\pm0.147)$~GeV~\cite{Yao:2015qia},  $\mathring{m}=(m_D+m_{D_s})/2$, $m_\ast=m_{D^\ast}$ for $(0,1/2)$ case and $m_\ast=m_{D_s^\ast}$ for $(1,0)$ case. The values for the physical masses are taken from PDG~\cite{Olive:2016xmw}. For the decay constants in the chiral limit, we take
\bea
{f_P}/{F_0}\to(1+\delta_\chi){f_D}/{F_\pi}\ ,
\eea
where the undetermined $\delta_\chi$ parameter accounts for higher-order chiral corrections.  Nonetheless, in the terms with $\beta_1$ and $\beta_2$, the $F_0$ is directly set to $F_\pi$.  Furthermore, here we use $F_\pi=92.4$~MeV~\cite{Olive:2016xmw} and $f_D=208.7$~MeV~\cite{Aoki:2016frl}. Finally, there is a total of four unknown parameters: $\beta_1$, $\beta_2$, $\beta_0$ and $\delta_\chi$, which we fit to LQCD data.

Best-fit parameters are compiled in Table~\ref{tab:fitsinD}, and the resulting scalar form factors are shown in Fig.~\ref{fig:charmfit}. The inner bands (statistical error) are obtained by varying the
parameters within their 1-$\sigma$ uncertainties in Table~\ref{tab:fitsinD}. The outer bands stand for the
total error by adding in quadratrue the systematic error from the MO matrix induced by the uncertainties of the LECs appearing in $\mathbf{T}^U$.  As one can see from the figure, the lattice data for $D\to\pi$ and
$D\to \bar{K}$ by the ETM Collaboration~\cite{Lubicz:2017syv} are well described in the whole kinematical region. The results of the HPQCD Collaboration~\cite{Na:2010uf,Na:2011mc} are also shown for comparison. Our predictions disagree with the HPQCD data in the region close to $q_{\rm max}^2$ for the $D\to\pi$ case. In Fig.~\ref{fig:charmfit}, we also predict the scalar form factors for the $D\to \eta$, $D_s\to K$ and $D_s\to \eta$
transitions. Predictions for the modulus of the vector form factors at $q^2=0$ are given in Table~\ref{tab:prediction}, where the statistical and systematical errors are shown in the first and second brackets respectively.
Using the experimental values~\cite{Amhis:2016xyh}, 
$
f_+^{D\to\pi}(0)|V_{cd}|=0.1426(19)$ and 
 $f_+^{D\to\bar{K}}(0)|V_{cs}|=0.722(5)
$, 
we obtain our predictions for the following CKM
elements:
\bea
|V_{cd}|=0.243(_{-12}^{+11})_{\rm sta.}({3})_{\rm sys.}(3)_{\rm exp.}\ ,\qquad |V_{cs}|=0.950(_{-40}^{+39})_{\rm sta.}(1)_{\rm sys.}(7)_{\rm exp.}\ .
\eea

\begin{table}[t]
\vspace{-0.5cm}
\begin{lrbox}{\leftbox}
\addtolength{\tabcolsep}{2pt}%
\begin{tabular}{lrcccc}
\toprule
 &{\rm Value}&\multicolumn{4}{c}{{\rm Correlation~matrix}}\\
 \cline{3-6}
$\chi^2$&1.82& $\beta_1$ & $\beta_2$  & $\beta_0$ &$\delta_\chi$\\
\midrule
$\beta_1$&$0.08(2)$&$1$&$0.96$&$-0.03$&$0.68$\\
$\beta_2$&$0.07(1)$&&$1$&$0.09$&$0.84$\\
$\beta_0$&$0.12(1)$&&&$1$&$0.11$\\
$\delta_{\chi}$&$-0.21(2)$&&&&$1$\\
\bottomrule
\end{tabular}
\end{lrbox}
\begin{lrbox}{\rightbox}
\begin{tabular}{lc}
\toprule
$|f_+(0)|$ &{\rm Value} \\ 
\midrule
$D\to\pi$&$0.587(_{-26}^{+28})(_{-8}^{+7})$\\
$D\to\eta$&$0.084(18)(_{-20}^{+29})$\\
$D_s\to K$&$0.425(_{-38}^{+36})(_{-36}^{+50})$\\
\midrule
$D\to \bar{K}$&$0.760(_{-31}^{+32})(1)~$\\
$D_s\to\eta$&$0.581(_{-20}^{+21})(_{-21}^{+19})$\\
\bottomrule
\end{tabular}
\end{lrbox}
\centering
\makebox[0pt]{%
\begin{minipage}[b]{1.2\wd\leftbox} 
\centering
\caption{\bf Fit results. }\label{tab:fitsinD}
\vspace{0.2cm}
\usebox{\leftbox}
\end{minipage}\quad
\begin{minipage}[b]{1.2\wd\rightbox}
\centering
\caption{\bf Predictions }\label{tab:prediction}
\vspace{0.2cm}
\usebox{\rightbox}
\end{minipage}%
}
\end{table}

\section{Summary}
The scalar form factors in the semileptonic heavy $D$ meson decays $D\to\pi\bar{\ell}\nu_\ell$ and $D\to\bar{K}\bar{\ell}\nu_\ell$ have been studied using the MO formalism. Coupled-channel effects, due to rescattering of the $P\phi$ system, are taken into account by solving coupled integral MO equations. The coefficients in the polynomials are constrained by light-quark chiral SU(3) symmetry. We fit the unknown parameters to the latest LQCD data  by ETM collaboration. The LQCD data are well described and the scalar form factors which are in the same multiplets as $D\to\pi(\bar{K})$ form factors are predicted in their kinematical regions. The CKM elements $|V_{cd}|$ and $|V_{cs}|$ are extracted as well. The extension to bottom sector is ongoing in a follow-up work~\cite{yao}.

\noindent{\bf Acknowledgements:}
This
research is supported by the Spanish Ministerio de Econom\'ia y Competitividad
and the European Regional Development Fund, under contracts FIS2014-51948-C2-1-P
and SEV-2014-0398, by Generalitat Valenciana under contract
PROMETEOII/2014/0068, by the National Natural Science Foundation of China
(NSFC) under Grant No.~11647601, by DFG and NSFC though funds provided to the
Sino-German CRC 110 ``Symmetries and the Emergence of Structure in QCD'' (NSFC
Grant No.
11261130311),  by the Thousand Talents Plan for Young Professionals, and by the
CAS Key Research Program of Frontier Sciences under Grant No.~QYZDB-SSW-SYS013.

\end{document}